\documentclass[%
 reprint,
superscriptaddress,
nolongbibliography,
 amsmath,amssymb,
 aps,longbibliography,
floatfix,
]{revtex4-2}

\usepackage[utf8]{inputenc}
\usepackage{color}
\usepackage{graphicx}
\usepackage{dcolumn}
\usepackage{bm}
\usepackage[colorlinks= true,urlcolor=blue,linkcolor= blue,citecolor=blue,bookmarks=false,pdfstartview=]{hyperref}
\usepackage{wasysym}
\usepackage{amstext}
\usepackage{gensymb}
\usepackage{float}
\usepackage{amssymb}
\usepackage{orcidlink} 
\usepackage{soul}
\usepackage{color}

\begin{document}

\preprint{APS/123-QED}
\title{Gapped 1/9 Magnetization Plateau in the Anisotropic Kagome Antiferromagnet Y-kapellasite}

\author{Dipranjan Chatterjee}
\email{dipranjan.chatterjee@physics.ox.ac.uk}
\affiliation{Universit\'{e}  Paris-Saclay,  CNRS,  Laboratoire  de  Physique  des  Solides,  91405,  Orsay,  France}
\affiliation{Clarendon Laboratory, Department of Physics, University of Oxford, Parks Road, OX1 3PU, United Kingdom}

\author{Paul A. Goddard}
\affiliation{Department of Physics, University of Warwick, Gibbet Hill Road, Coventry, CV4 7AL, UK}

\author{Ewan R. P. Thomas}
\affiliation{Clarendon Laboratory, Department of Physics, University of Oxford, Parks Road, OX1 3PU, United Kingdom}%

\author{Katharina M. Zoch}
\affiliation{Physikalisches Institut, Goethe-Universit\"at Frankfurt, Frankfurt am Main, Germany}

\author{Hank C. H. Wu}
\affiliation{Clarendon Laboratory, Department of Physics, University of Oxford, Parks Road, OX1 3PU, United Kingdom}%

\author{Benjamin M. Huddart}
\affiliation{Clarendon Laboratory, Department of Physics, University of Oxford, Parks Road, OX1 3PU, United Kingdom}%
\author{Cornelius Krellner}
\affiliation{Physikalisches Institut, Goethe-Universit\"at Frankfurt, Frankfurt am Main, Germany}

\author{Edwin Kermarrec}
\affiliation{Universit\'{e}  Paris-Saclay,  CNRS,  Laboratoire  de  Physique  des  Solides,  91405,  Orsay,  France}%

\author{Mladen Horvati{\'c}\,\orcidlink{0000-0001-7161-0488}}
\affiliation{
Laboratoire National des Champs Magn\'{e}tiques Intenses, LNCMI-CNRS (UPR3228), EMFL, \\
Univ. Grenoble Alpes, Univ. Toulouse, INSA-T, 38042 Grenoble Cedex 9, France}

\author{Steffen Kr\"{a}mer\,\orcidlink{0000-0002-6107-3583}}
\affiliation{
Laboratoire National des Champs Magn\'{e}tiques Intenses, LNCMI-CNRS (UPR3228), EMFL, \\
Univ. Grenoble Alpes, Univ. Toulouse, INSA-T, 38042 Grenoble Cedex 9, France}

\author{Pascal Puphal}
\affiliation{Max-Planck-Institute for Solid State Research, Heisenbergstra{\ss}e 1, 70569 Stuttgart, Germany}
\affiliation{1.~Physikalisches Institut, Universit\"at Stuttgart, Pfaffenwaldring 57, 70569 Stuttgart, Germany}%

\author{John Singleton}
\affiliation{National High Magnetic Field Laboratory (NHMFL), Los Alamos National Laboratory, Los Alamos, NM, USA}

\author{Stephen J. Blundell}
\affiliation{Clarendon Laboratory, Department of Physics, University of Oxford, Parks Road, OX1 3PU, United Kingdom}%

\author{Fabrice Bert}
\affiliation{Universit\'{e}  Paris-Saclay,  CNRS,  Laboratoire  de  Physique  des  Solides,  91405,  Orsay,  France}%

\date{\today}%

\begin{abstract}

Fractional magnetization plateaus provide a sensitive probe of many-body spin states in frustrated quantum magnets, yet their microscopic origin in kagome antiferromagnets remains unresolved. This is particularly true of the mysterious $1/9$ plateau, which is predicted by theory but infrequently observed in experiment. Here, we investigate this problem in the $S = 1/2$ anisotropic kagome antiferromagnet Y-kapellasite, Y$_3$Cu$_9$(OH)$_{19}$Cl$_8$, using pulsed-field magnetization measurements on single crystals and high-field $^{35}$Cl NMR. We identify a hierarchy of field-induced fractional features, including $1/3$ and $1/9$ plateaus, as well as a weaker low-field feature. Analysis of the NMR spectra and the magnetic susceptibility across the $1/9$ plateau demonstrate that it is accompanied by an ordered local spin configuration, a strong suppression of low-energy spin fluctuations and activated behavior, consistent with a gapped fractional state. These features differ from those in the only other material YCu$_3$(OH)$_6$Br$_2$[Br$_{1-y}$(OH)$_y$] in which this plateau is observed, implying a surprising robustness of the $1/9$ state to the details of the underlying magnetism.  
\end{abstract}

\maketitle

\footnotetext[1]{See Supplemental Material at [URL will be inserted by publisher] for further details, which includes Refs. \cite{khasanov2016high,Khasanov2022,mezouar2024high,garbarino2024extreme,Kremer2025,Boldrin2015,Chatterjee2023,Kermarrec2014,Khuntia2020,Dolezal2024,Celeste2019,Gruneisen_1912}, for additional information.}

The search for an unambiguous realization of a spin liquid continues. In $S = 1/2$ Heisenberg kagome antiferromagnets, a key indicator of spin-liquid physics is expected to be the presence of a $1/9$ magnetization plateau~\cite{yoshida2022frustrated,He2024,nishimoto2013controlling, Picot}. So far it has been observed in only one real material~\cite{Zeng2022,zheng2025unconventional} and further experimental evidence is required to determine the universality of this $1/9$ feature. Progress is limited by both the scarcity of suitable kagome materials and the very large magnetic fields typically needed to access the relevant regime where magnetization plateaus can be observed~{\cite{Haraguchi,Ono,Yoshii,ishikawa, Chen, Ishii, ueda,herbertsmithite,yoshida2022frustrated,okuma2019series,nishimoto2013controlling, Picot}.

Kapellasite-structured \cite{Kermarrec2014,Fak12, Bernu, Messio} kagome antiferromagnets provide a tunable platform for plateau physics, where competing interactions can drive the system between ordered and fluctuation-dominated regimes by modest structural changes. In In-kapellasite~\cite{kato2025magnetic,kato2024one}, the one-third plateau appears together with a sizable field-induced temperature-linear heat-capacity term, consistent with low-energy gapless excitations nearby. In the related Y-based kagome compound $\mathrm{YCu_3(OH)_6Br_2[Br_{1-y}(OH)_y]}$ (YCOB), high-field magnetization reveals not only a $1/3$ feature, but also a clear $1/9$ plateau~\cite{Zeng2022,zheng2025unconventional}, previously seen only in theoretical calculations~\cite{nishimoto2013controlling,Picot,CHEN2018}. Torque magnetometry additionally detects oscillations in this Mott insulator, emerging above the $1/9$ field scale~\cite{zheng2025unconventional}. These observations, as well as heat capacity data~\cite{Zeng2022,ZhengHeatCap2025}, have been interpreted within a phenomenological Dirac-spinon picture, in which the applied field drives the spinon chemical potential through a Dirac node, generating an effective gauge flux and Landau-level-like quantization. The resulting V-shaped $\mathrm{d}M/\mathrm{d}H$ minimum and its non-saturating value at low-temperature  have been taken as signatures of gapless fermionic excitations associated with an unconventional $1/9$ plateau state. Nevertheless, theoretical consensus is difficult to reach while experimental data on the $1/9$ plateau remain sparse.

\begin{figure*}[t]
\begin{centering}
\includegraphics[width=2\columnwidth]{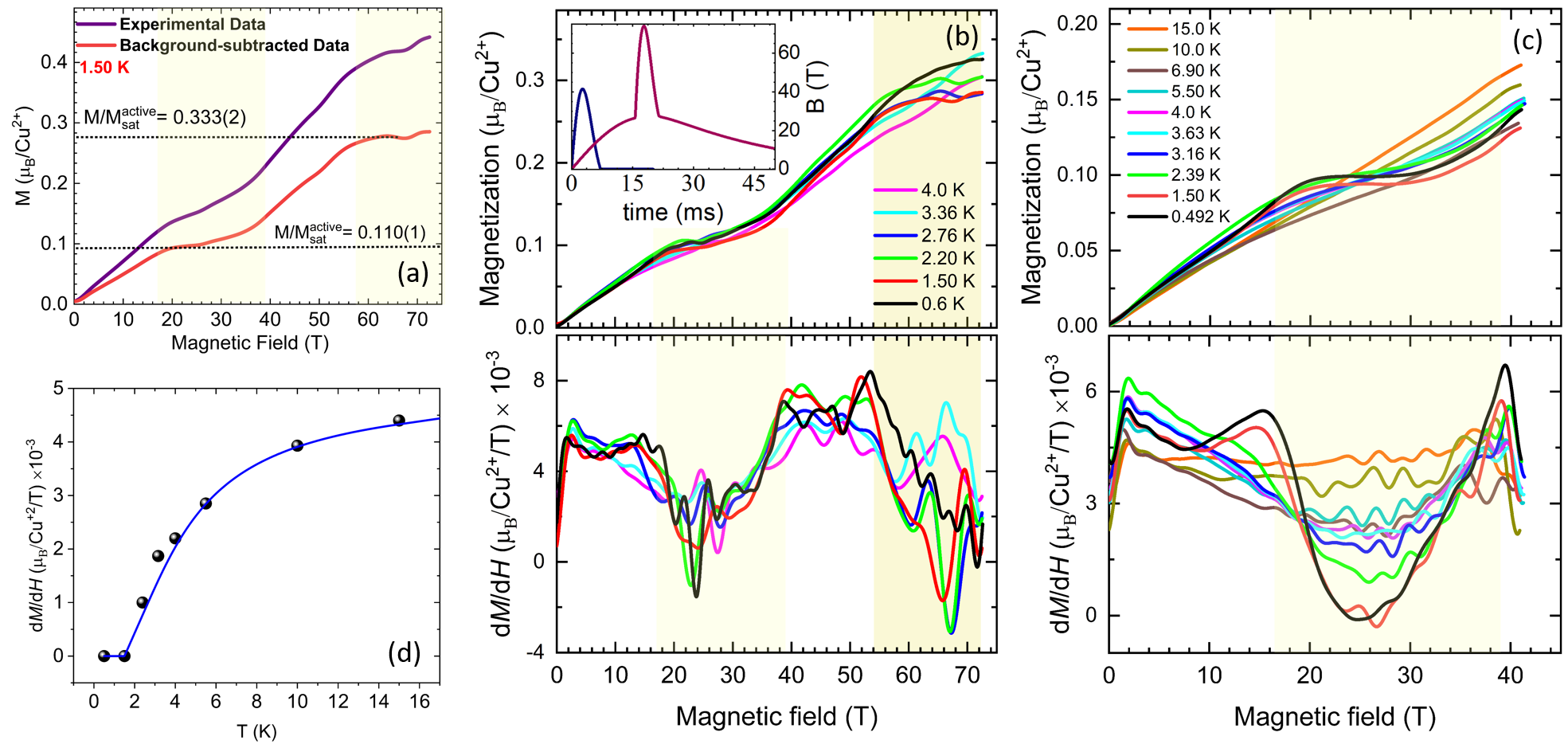}
\par\end{centering}
\caption{\textbf{Field-induced fractional magnetization plateaus in Y-kapellasite ($B\parallel c$).} (a) Magnetization $M(H)$ at $T=1.5$ K showing the experimentally acquired data (purple) and background-subtracted data (red), obtained by removing a linear contribution attributed to $30\%$ of Cu moments. Normalizing the subtracted magnetization to the expected saturation value of the $70\%$ plateau-active spins, reveals plateaus at $M/M_{\rm sat}^{\rm active} \simeq 0.110(1)$ and $0.333(2)$, consistent with $1/9$ and $1/3$ fractions. Yellow shading throughout indicates the predicted $1/9$ and $1/3$ plateau field windows for a uniform exchange energy of 112~K~\cite{Raikos2026}. (b) Background-subtracted magnetization $M(H)$ and differential susceptibility $\mathrm{d}M/\mathrm{d}H$ measured up to 75~T at Los Alamos National Laboratory for different temperatures. The inset shows $B(t)$ profiles for the slower LANL and faster Oxford field pulses. The $\mathrm{d}M/\mathrm{d}H$ data show a pronounced suppression across the $1/9$ plateau and a second high-field suppression consistent with the $1/3$ plateau. (c) Magnetization close to the $1/9$ plateau performed at the Nicholas Kurti Magnetic Field Laboratory (Oxford). The $1/9$ feature broadens and vanishes by $T\sim15$~K. (d) Temperature dependence of the minimum in $\mathrm{d}M/\mathrm{d}H$ at the $1/9$ plateau (points). The evolution follows a Fermi–Dirac thermal broadening model (line) \cite{zheng2025unconventional} above the ordering temperature.}
 \label{plat}
\end{figure*}

Y-kapellasite, Y$_3$Cu$_9$(OH)$_{19}$Cl$_8$ \cite{Hering2022,Chatterjee2023,Biesner22,Chatterjee2026}, offers a complementary and well-constrained setting. Unlike YCOB, which hosts a nonmagnetic ground state with substantial exchange disorder, Y-kapellasite has an ordered bond-modulated kagome network, an enlarged in-plane superstructure, and zero-field long-range magnetic order with a reduced ordered moment~\cite{Chatterjee2023}. This contrast sharpens the central question: can a $1/9$ fractional plateau also emerge from an anisotropic kagome magnet with pre-existing magnetic order, and does the resulting field-induced state remain gapless? Addressing this requires systematic high-field probes on a clean and structurally well-characterized platform, which motivates the present study. 

In this Letter, we report pulsed-field magnetization measurements on single crystals of Y-kapellasite up to 75~T. These measurements provide direct access to the field-induced magnetization process and allow fractional plateau regimes to be resolved. To establish the microscopic character of the plateau state, we complement the bulk magnetization data with high-field $^{35}$Cl NMR measurements, which probe the local static spin configuration and low-energy spin dynamics.

Prior NMR, $\mu$SR and neutron scattering measurements have shown that Y-kapellasite exhibits $(1/3, 1/3)$ magnetic order below $T_{\rm N} = 2.1$~K  and has a regular arrangement of three nearest-neighbor antiferromagnetic exchange bonds ($J = 140, 140, 56$~K)~\cite{Chatterjee2023,Wang2023}. Unidentified features in the magnetization were also previously reported in the Supplementary Information of Ref.~\cite{Biesner22}. 

The results of our 75~T magnetization measurements performed at the National High Magnetic Field Laboratory in Los Alamos are shown in Fig.~\ref{plat}(a) and (b). For $B\parallel c$, the magnetization is initially linear at low fields, before revealing a clear sequence of field-induced plateau-like features.  Data for $B \perp c$ are presented in the \textcolor{black}{End Matter} and are nearly identical, consistent with the weak magnetic anisotropy of Y-kapellasite~\cite{Chatterjee2023}.
The experimentally acquired $M(H)$ curves retain a finite slope even within the plateau-like regions, which is found to be approximately $30\%$ of the low-field $M(H)$ gradient, suggesting the presence of a linear background contribution in addition to the fractional plateau response that arises from the remaining $70\%$ of the spins. In this case, the plateau-active $70\%$ spin fraction would yield a saturated moment $M_{\rm sat}^{\rm active} \simeq 0.84(1)~\mu_{\rm B}/{\rm Cu}^{2+}$, using the previously determined value of $g_c \simeq 2.40(1)$ \cite{Chatterjee2023}. Subtracting $30\%$ of the low-field linear susceptibility from the $M(H)$ data at each temperature reveals plateaus occurring at $M/M_{\rm sat}^{\rm active} \simeq 0.110(3)$ and $M/M_{\rm sat}^{\rm active} \simeq 0.33(1)$, consistent with $1/9$ and $1/3$ plateau values of the active spin contribution. 

After subtraction, both plateaus appear as pronounced dips in $\mathrm{d}M/\mathrm{d}H$ centered at $\mu_0 H = 25.6(3)$~T and $65.5(5)$~T, identifying the $1/9$ and $1/3$ plateau fields, respectively (Fig.~\ref{plat}(b)). The expected field positions of the fractional plateaus of the ideal $S = 1/2$ kagome Heisenberg model are well modeled experimentally in terms of the dimensionless ratio $h/J$, where $h = g\mu_B B$ is the Zeeman energy and $J$ is the nearest-neighbor exchange~\cite{nishimoto2013controlling,CHEN2018,Raikos2026}. The shaded regions in Fig.~\ref{plat}(b) and (c) show the expected widths in field of the $1/9$ and $1/3$ plateau regions derived using the variational wavefunctions approach of Ref.~\cite{Raikos2026} and $J/k_{\rm B} = 112$~K, the average of the three anisotropic exchange energies in Y-kapellasite as determined from neutron scattering measurements~\cite{Chatterjee2023}. The theoretical predictions agree extremely well with the location of the dips in $\mathrm{d}M/\mathrm{d}H$, corroborating their identification as fractional kagome plateau states. 

The 75~T duplex magnet used in Los Alamos has a sharp change in $\mathrm{d}B/\mathrm{d}t$ at 26~T where the inner coil activates (see inset to Fig.~\ref{plat}(b)), which coincides with the $1/9$ plateau. To avoid this and to perform a systematic temperature-dependent study of the $1/9$ region, we performed additional measurements up to 42~T using a single-coil magnet in the Nicholas Kurti Magnetic Field Laboratory in Oxford. The resulting background-subtracted data are shown in Fig.~\ref{plat}(c) with the $1/9$ plateau clearly observed. In both datasets, an additional weaker low-field feature is visible as a shallow depression in $\mathrm{d}M/\mathrm{d}H$ at $\mu_0 H =  7.90(5)$~T, occurring at $M/M_{\rm sat}^{\rm active} = 0.0475$\textcolor{black}{$(3)\approx 1/21$}, although it is not resolved as a fully formed plateau in $M(H)$ down to 0.49~K, the lowest temperature achieved.

As temperature is increased, the minima in $\mathrm{d}M/\mathrm{d}H$ at all plateaus broaden and lose contrast. The low-field depression is discernible only below 2~K, while the 1/3 plateau is largely gone by 4~K. By contrast, the $1/9$ is still apparent up to about 10~K. Similar to YCOB, the feature at our $1/9$ plateau appears as a V-shaped dip in $\mathrm{d}M/\mathrm{d}H$, deepening swiftly on cooling. The depth of the feature saturates at low temperatures in our case, as shown in Fig.~\ref{plat}(d). This is in contrast to YCOB, where the dip continues to deepen on cooling, which was given as evidence of a gapless excitation spectrum in this field region~\cite{zheng2025unconventional}. Following the analysis of YCOB presented in Ref.~\cite{zheng2025unconventional}, we find that the temperature dependence of the depth of the $1/9$ dip is well described by a Fermi-Dirac thermal broadening model for temperatures above 1.5~K (see Equation~\ref{FDmodel} in the End Matter). This is consistent with the idea that the $1/9$ plateau is linked to a Fermi surface of Dirac spinons.

\begin{figure}[!]
\begin{centering}
\includegraphics[width=1\columnwidth]{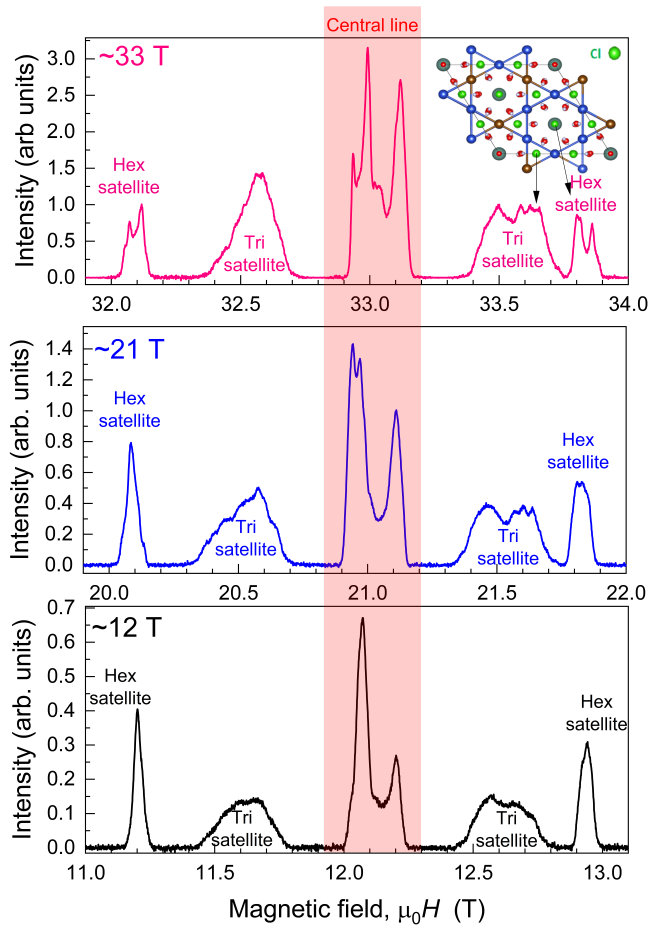}
\par\end{centering}
\caption{\textbf{Field-swept $^{35}$Cl NMR spectra of Y-kapellasite at 1.2~K near $\mu_0H=$ 12, 21, and 33 T.}
The shaded region marks the central transition $(m_I=+1/2 \leftrightarrow -1/2)$, while the outer features are the quadrupolar satellites $(m_I=\pm 1/2 \leftrightarrow \pm 3/2)$. Two distinct satellite components are resolved and assigned to the Tri and Hex $^{35}$Cl sites, corresponding to Cl nuclei in inequivalent local environments of the kagome network. Inset: schematic of the crystal structure showing the two site assignments.}
 \label{Cl35}
\end{figure}

High-field $^{35}$Cl NMR measurements were carried out at LNCMI Grenoble. Characteristic spectra obtained at 1.2~K in the field range $12$--$33$~T are presented in Fig~\ref{Cl35}. Both the central ($-1/2 \leftrightarrow 1/2$) and satellite ($\pm 3/2 \leftrightarrow \pm 1/2$) transitions of $^{35}$Cl with nuclear spin $I=3/2$ are resolved and assigned following Ref.~\cite{Chatterjee2023} to the two Cl crystallographic sites, respectively located at the centers of the Cu hexagon and Cu triangles (labeled 'Hex' and 'Tri' in the inset of Fig.~\ref{Cl35}). The broad and complex central line is overall consistent with the spectra reported below $T_\mathrm{N} = 2.1$~K at $\mu_0 H \simeq 7.7$~T~\cite{Chatterjee2023}. It reflects the overlap of the two Cl site contributions, each with a different internal field distribution. Namely the Tri Cl site yields three overlapping broad lines in the ordered phase, extending on both side of the Hex site contribution, which remains comparatively narrower. With increasing fields up to the 1/9 plateau phase (Fig.~\ref{Cl35} and Fig.~\ref{NMLine}), the central lineshape evolves smoothly as a result of the different hyperfine couplings of the two lines and smooth changes in the spin texture. Remarkably, the lineshape becomes nearly field independent in the plateau regime, showing that the constant magnetization corresponds to a locking of the magnetic structure. At still higher fields, beyond the 1/9 plateau, the lineshape evolves again. At 33~T a splitting of the Hex line, best observed on the corresponding satellite lines (Fig.~\ref{Cl35} top panel), signals a qualitatively different magnetic phase. To quantify the evolution of the central line, we evaluate its second moment $M_2$. It is obtained from the intensity-weighted variance of the field distribution, and the linewidth is taken as $\sigma=\sqrt{M_2}$. As shown in Fig.~\ref{Cl35}(d), $\sigma$ remains nearly field independent across the plateau regime, indicating that the distribution of local hyperfine fields is essentially unchanged. This behavior points to a stable microscopic spin configuration, with no significant redistribution of local magnetic fields over this field range.

\begin{figure}[hbt!]
\begin{centering}
\includegraphics[width=1\columnwidth]{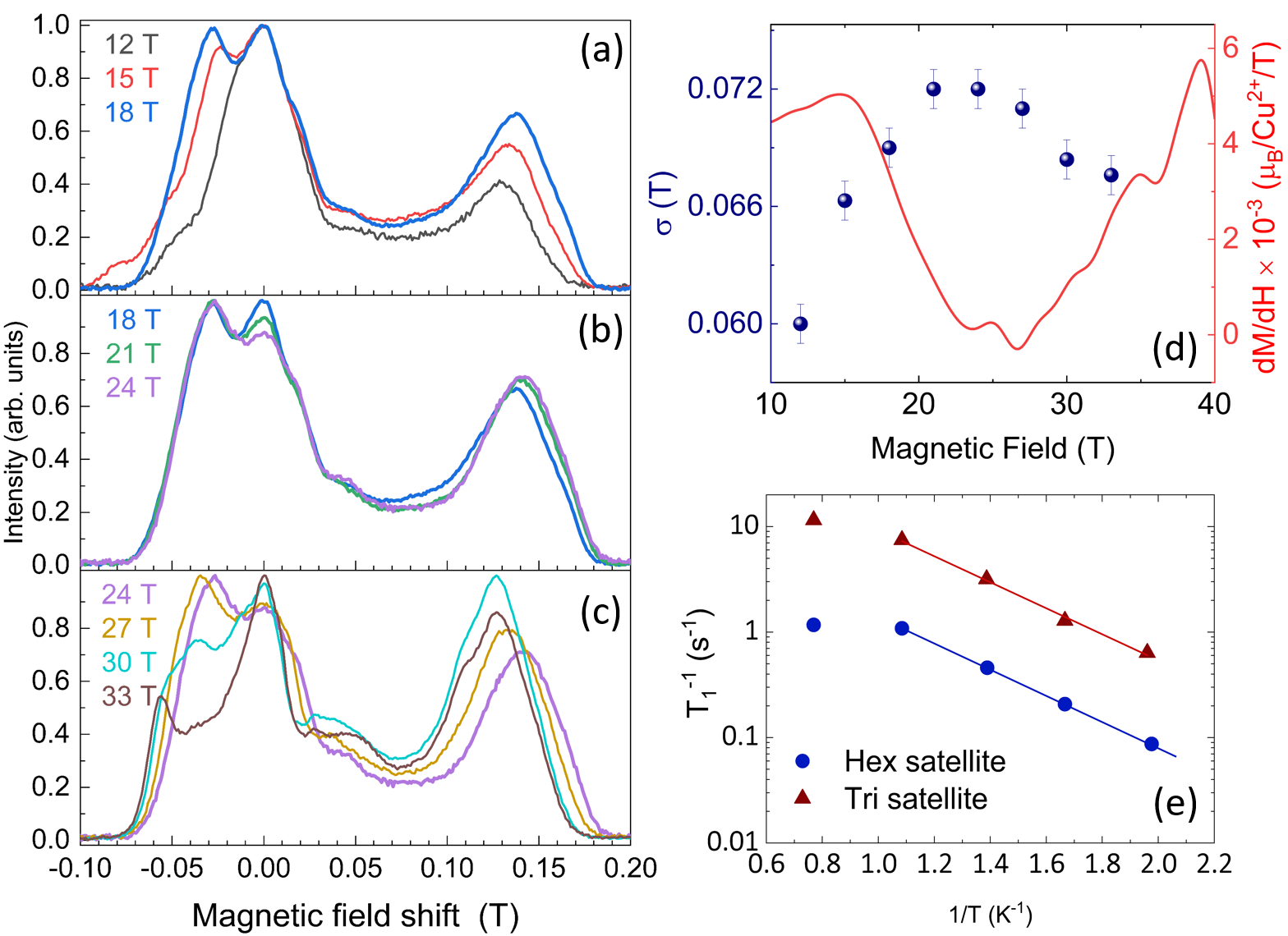}
\par\end{centering}
\caption{\textbf{$^{35}$Cl NMR response across the $1/9$ plateau regime in Y-kapellasite.}
(a–c) Field-swept $^{35}$Cl NMR spectra at selected magnetic fields at $T = 1.2$~K. (d) Field dependence of the NMR second moment, $\sigma = \sqrt{M_2}$ at $T = 1.2\,\mathrm{K}$ (left axis), together with the differential susceptibility $\mathrm{d}M/\mathrm{d}H$ at $T = 1.5\,\mathrm{K}$ (right axis).  The suppression of $\mathrm{d}M/\mathrm{d}H$ across the plateau region, accompanied by an essentially field-independent $\sigma$, indicates a stabilization of the local hyperfine field distribution. (e) Temperature dependence of $T_1^{-1}$ for the hexagonal and triangular satellites are measured at $\mu_0H \simeq 21$~T, showing activated behavior consistent with a gapped spin state in the plateau regime.
}
\label{NMLine}
\end{figure}

Further, we probe the spin dynamics in the 1/9 plateau phase by measuring the temperature dependence of the NMR spin-lattice relaxation time. To avoid the mixing of the two Cl site contributions at the central line position, $T_1$ was measured separately on the two resolved NMR satellites, located at $\mu_0H = 20.08$~T for the hexagonal site and $\mu_0H = 20.57$~T for the triangular site. As shown in 
Fig.~\ref{NMLine}(e), $\log(1/T_1)$ varies linearly with $1/T$ for both sites, consistent with activated spin fluctuations. The nearly parallel slopes indicate a common activation scale, while the difference in absolute magnitude reflects the distinct hyperfine couplings at the two sites. From the fitted slope, we extract an activation energy $\Delta/k_{\mathrm B}\approx 2.8(2)$~K, roughly  compatible with $\mathrm{d}M/\mathrm{d}H$ being saturated below 2 K [Fig.~\ref{plat}(d)].
These results show that the $1/9$ plateau-like state is associated with a spin-gapped low-energy excitation spectrum.

Our study of Y-kapellasite identifies well-defined $1/9$ and $1/3$ fractional magnetization plateaus in this anisotropic kagome antiferromagnet. Moreover, it is apparent that only approximately $70\%$ of the spins participate in the plateau states, while the remaining $30\%$ form a weakly field-dependent background. The origin of this partition is not clear, but an important clue may be that the underlying bond-modulated kagome lattice of Y-kapellasite consists of approximately $30\%$ isotropic hexagons and $70\%$ distorted hexagons built from three inequivalent exchange paths~\cite{Hering2022,Chatterjee2023}. A similar $70/30$ structural motif has been reported in YCOB~\cite{Liu}, but here the distorted and undistorted hexagons are randomly distributed, while in Y-kapellasite they form an ordered superstructure. Further investigations are needed to establish whether this structural difference 
can account for the portion of the moment that contributes to the plateau states in Y-kapellasite.  

What is clear from our data on Y-kapellasite, is that the appearance of a $1/9$ plateau extends beyond the experimental framework provided by YCOB. There are similarities between the data in Y-kapellasite and YCOB. In both cases, the $1/9$ plateau appears as a V-shaped dip in $\mathrm{d}M/\mathrm{d}H$ that deepens on cooling, which has been taken to indicate spinon excitations with a Dirac-like dispersion~\cite{zheng2025unconventional}. But there are also differences between the two materials. In YCOB, the dip in $\mathrm{d}M/\mathrm{d}H$ continues to grow linearly at the lowest temperatures, which was attributed to gapless spinon modes. By contrast in Y-kapellasite, the dip bottoms out at 1.5~K and the NMR data clearly show the presence of a spin-gap in the excitation spectrum 
within the $1/9$ field regime. Moreover, YCOB remains magnetically disordered down to the lowest temperatures measured, while the zero-field ground state of Y-kapellasite exhibits magnetic order below $T_{\rm N} = 2.1$~K~\cite{Chatterjee2023}, 
although this order was shown to be fragile, easily suppressed by a moderate external pressure~\cite{Chatterjee2026}.
Our observed low-field feature centred at 7.90(5)~T is seen only at temperatures below $T_{\rm N}$, but the $1/9$ plateau appears disconnected from the magnetic order; it progressively broadens with increasing temperature and remains observable up to $ 10$~K, well above $T_N$. 

In the ideal $S = 1/2$  Heisenberg kagome model, the $1/9$ plateau has been proposed to be an unconventional state, possibly a $\mathbb{Z}_3$ spin liquid, distinct from the symmetry-broken plateau states that appear at higher fields~\cite{yoshida2022frustrated,He2024,nishimoto2013controlling, Picot}. The appearance of quantum oscillations in magnetic torque measured in YCOB, which — like Y-kapellasite — is an insulator, is given as evidence of a Fermi surface of chargeless Dirac spinons~\cite{zheng2025unconventional}. The $1/9$ plateau is interpreted within this picture, resulting from the existence of nine spinon bands arising from a tripling of the unit cell, caused either by broken translational symmetry or as a natural outcome of the $\mathbb{Z}_3$ state~\cite{zheng2025unconventional}. The present results on Y-kapellasite are consistent with this picture insofar as the susceptibility at the plateau has the same V-shaped dip and Fermi-Dirac temperature dependence (once the temperature exceeds 2~K).  However, our results also suggest that the $1/9$ plateau is impervious to certain aspects of the underlying magnetism. Y-kapellasite departs from both the ideal kagome Heisenberg framework and the experimental architecture of YCOB, and hosts a $1/9$ plateau in a system that incorporates a well-defined arrangement of exchange-bond anisotropy, proximity to antiferromagnetic order, and a gapped local excitation spectrum. Taken together, the results raise the possibility that the 70\% portion of the spins which are contributing to the magnetization plateaus coexist with the remaining 30\% that give rise to the low-field feature, the observed low-moment magnetic order and the excitation gap.

Finally, we note that Y-kapellasite is a particularly promising system in which to advance the experimental investigation of $1/9$-plateau physics, because its degree of geometric frustration, and thus the balance of competing interactions, can be controlled continuously by pressure~\cite{Chatterjee2026}, providing a direct means to tune the fractional plateau states.

\maketitle

\section*{Acknowledgments}
 DC, HCHW, BMH, ERPT, and SJB acknowledge funding from UK Research and Innovation (UKRI) through the UK government’s Horizon Europe funding guarantee (Grant No.~EP/X025861/1). The pulsed magnet
system as the Nicholas Kurti High Magnetic Field
Laboratory was refurbished with a grant from the UK
Engineering and Physical Sciences Research Council
(EPSRC) [Grant No. EP/J013501/1]. We thank Simon Davila Solano and Anthony Hickman for technical assistance.
A portion of this work
was performed at the National High Magnetic Field
Laboratory (NHMFL), which is supported by National
Science Foundation Cooperative Agreements No. DMR-
1644779 and No. DMR-2128556 and the Department of
Energy (DOE). 
FB and EK acknowledge the support of the French Agence Nationale de la Recherche, under Grant No. ANR-25-CE30-2010-01 “ULTIMAT” and of the Fondation Charles Defforey-Institut de France. This work was supported by LNCMI-CNRS, a member of the European Magnetic Field Laboratory (EMFL). The authors would like to acknowledge valuable discussions with Prof. Philippe Mendels.

\sloppypar
\bibliography{library.bib}
\onecolumngrid
\section*{End matter}
\twocolumngrid
\textbf{Magnetization data $B \perp c :$} To complement the $B \parallel c$ magnetization data, we also present measurements for $B \perp c$ in Fig.~\ref{plat-1}. Similar $1/9$ and $1/3$ magnetization plateaus are observed in this field orientation.

\begin{figure}[h!]
\begin{centering}
\includegraphics[width=1\columnwidth]{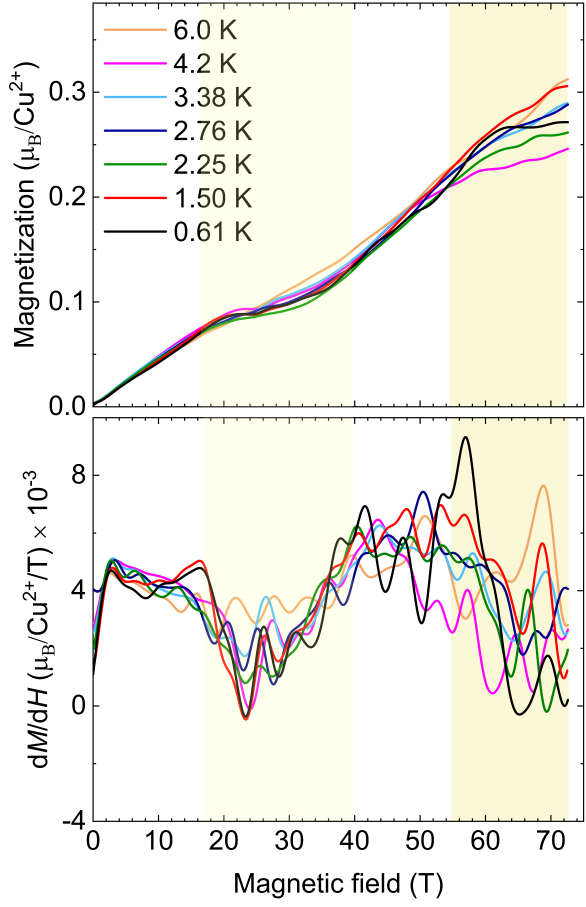}
\par\end{centering}
\caption{\textbf{Field-induced fractional magnetization plateaus in Y-kapellasite ($B \perp c$).}
Top panel: Background-subtracted magnetization $M(H)$ measured up to 75~T at Los Alamos National Laboratory for several temperatures. Yellow shaded regions indicate the predicted field windows for the $1/9$ and $1/3$ magnetization plateaus.
Bottom panel: Differential susceptibility $dM/dH$ as a function of magnetic field. The $dM/dH$ data show a pronounced suppression within the $1/9$ plateau region and a second high-field suppression consistent with the onset of the $1/3$ plateau.}
 \label{plat-1}
\end{figure}

\textbf{Fermi-Dirac model.} The temperature dependence of the  susceptibility data at the center of the $1/9$ plateau is fitted using the Fermi-Dirac thermal broadening model of Ref~\cite{zheng2025unconventional}, adapted to account for the low-temperature saturation at $T_0$ seen in Y-kapellasite:
\begin{equation}
    \chi(T) = \chi_\textit{p}\left[1+\frac{T-T_0}{T_\textit{p}}\ln\left(\frac{2}{1+\cosh\left(\frac{T_\textit{p}}{T-T_0}\right)}\right)\right]
    \label{FDmodel}
\end{equation}
The data in Fig.~\ref{plat}(d) are well described by this model above $T_0 = 1.5(4)$~K with the parameters $\chi_\textit{p} = 5.0(4)\times 10^{-3}\,\mu_\textit{B}$T$^{-1}$ per Cu$^{2+}$ and $T_\textit{p} = 8(2)$~K.

\end{document}